\documentclass[12pt,psfig]{article}
\usepackage{graphicx}
\begin{document}

\begin{center} 
{\bf  The neutron skin of $^{208}$Pb and the density dependence of the symmetry energy                   
 }              
\end{center}
\vspace{0.1cm} 
\begin{center} 
 Francesca Sammarruca and Pei Liu \\ 
\vspace{0.2cm} 
 Physics Department, University of Idaho, Moscow, ID 83844, U.S.A   
\end{center} 
\begin{abstract}
We explore neutron skin predictions for $^{208}$Pb in relation to the symmetry pressure 
in various microscopic models based on realistic nucleon-nucleon potentials and either the 
Dirac-Brueckner-Hartree-Fock approach or the 
conventional Brueckner-Hartree-Fock framework implemented with three-body forces. 
We also discuss the correlation between the neutron skin and the radius of a fixed-mass
neutron star. 
\\ \\ 
PACS number(s): 21.65.+f, 21.30.Fe 
\end{abstract}

\section{Introduction} 
                                                                     
Microscopic predictions of the nuclear equation of state (EoS), together with empirical constraints 
from EoS-sensitive observables, are a powerful combination to learn about the in-medium
behavior of the nuclear force. With this objective in mind, 
over the past several years, our group has taken a broad look at the EoS exploring diverse 
aspects and extreme states of nuclear matter.

From the experimental side, intense effort is going on to obtain reliable empirical information for the less
known aspects of the EoS. Heavy-ion (HI) reactions are a popular way to seek constraints on the symmetry 
energy, through analyses of observables that are sensitive to the pressure gradient between 
nuclear and neutron matter. 
Isospin diffusion data from HI collisions, together with analyses based on isospin-dependent transport
models, provide information on the slope of the symmetry energy. Naturally, different
reaction conditions, in terms of energy per nucleon and/or impact parameter, will probe different density regions. 

Concerning the lower densities,     
isospin-sensitive observables can also be identified among the properties of normal nuclei. 
The neutron skin of neutron-rich nuclei is a powerful isovector observable, being sensitive to the   
slope of the symmetry energy, which determines to which extent neutrons are
``pushed out" to form the skin. 
It is the purpose of this note to systematically examine and discuss the symmetry energy properties 
 in microscopic models and the corresponding neutron skin predictions.  We will take the skin of
$^{208}$Pb as our representative isovector ``observable".
                           
Parity-violating electron scattering experiments are now a realistic option        
to determine neutron distributions with unprecedented accuracy. The neutron radius of 
$^{208}$Pb is expected to be measured within 0.05 fm thanks to the electroweak program
at the Jefferson Laboratory \cite{Piek06}. This level of accuracy could not be achieved with hadronic scattering due to the large theoretical uncertainties present in all hadronic models. 
Parity-violating electron scattering at low momentum transfer is especially suitable to probe neutron densities, as the                   
 $Z^0$ boson couples primarily to neutrons. 
With the success of this program, 
 reliable empirical information on neutron skins will be able to provide, in turn, much needed {\it independent} constraint on the 
density dependence of the symmetry energy. 

Furthermore, 
with the Facility for Rare Isotope Beams (FRIB) recently approved for design and construction at 
Michigan State University, 
studies of neutron-rich systems become particularly important and timely. Such program will have widespread
impact, reaching from the physics of exotic nuclei to nuclear astrophysics.

\section{Predictions of neutron densities with a simple energy functional} 

We calculate proton and neutron density distributions with a method described in an earlier work \cite{AS03}. 
Namely, we use an energy functional based on the semi-empirical mass formula, where the volume and  
symmetry terms are contained in the isospin-asymmetric equation of state. Thus, we write the 
energy of a (spherical) nucleus as 
\begin{equation}
E(Z,A) = \int d^3 r~ e(\rho(r),\alpha(r))\rho(r) + 
\int d^3 r f_0(|\nabla \rho(r)|^2 + \beta 
|\nabla \rho_I(r)|^2) + Coulomb~ term. 
\end{equation} 
In the above equation, 
$\rho$ and $\rho_I$ are the usual isoscalar and isovector densities, given by, (in terms of the neutron and the proton densities), $\rho_n +\rho_p$ and 
$\rho_n -\rho_p$, respectively, $\alpha$ is the neutron asymmetry
parameter, $\alpha=\rho_I/\rho$, and $e(\rho,\alpha)$ is the energy per particle in 
isospin-asymmetric nuclear matter. 
We use 
realistic nucleon-nucleon forces, (specifically, Bonn B \cite{Mac89}), and 
our latest EoS from Ref.~\cite{Sam0806}. The latter is based on the Dirac-Brueckner-Hartree-Fock method, 
the technical framework of which was described earlier \cite{AS03I}.  
The energy per particle in 
symmetric and neutron matter are shown in Fig.~1. 
\begin{figure}[h]
\begin{center}
\vspace*{-4.0cm}
\hspace*{-2.0cm}
\scalebox{0.4}{\includegraphics{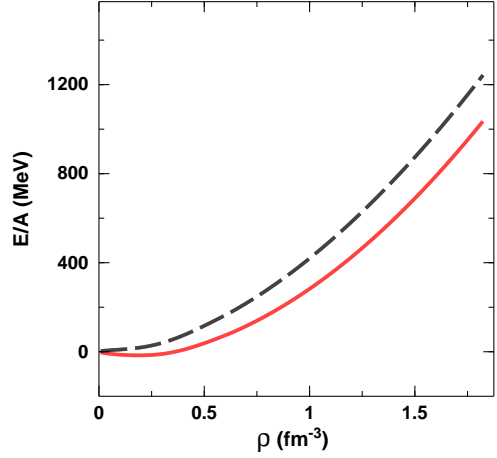}}
\vspace*{-3.0cm}
\caption{(color online) Energy/particle in symmetric matter (solid red) and neutron matter (dashed black)  as from 
our latest DBHF calculations. 
} 
\label{one}
\end{center}
\end{figure}

From fits to nuclear binding energies, 
the constant $f_0$ in Eq.~(1) is approximately 70 $MeV$ $fm^5$, whereas the contribution of the term proportional to $\beta$
was found to be minor \cite{Oya98}. Thus we will neglect it.           
(The magnitude of  $\beta$ was estimated to be about 1/4 in Ref.~\cite{Furn}, where it was observed that, 
even with variations of $\beta$ between -1 and +1, the effect of the $\beta$ term on the neutron skin 
was negligibly small.) 

The parameters of the proton and neutron density functions are obtained by minimizing the value
of the energy, Eq.~(1), assuming Thomas-Fermi distributions. 
Although simple, this method has the advantage of allowing a         
very direct connection between the EoS and the properties of finite nuclei. (It could be used, for 
instance, to determine a semi-phenomenological EoS by fitting to both binding energies and
charge radii of closed-shell nuclei.) 

In Fig.~2, we show the proton and neutron Thomas-Fermi distributions for $^{208}$Pb as obtained with the method describe above and the DBHF model for the       
EoS. The predicted proton and neutron root-mean-square radii are 
 5.39 fm and 5.56 fm, respectively.                                                                      
\begin{figure}[h]
\begin{center}
\vspace*{1.0cm}
\hspace*{-2.0cm}
\scalebox{0.3}{\includegraphics{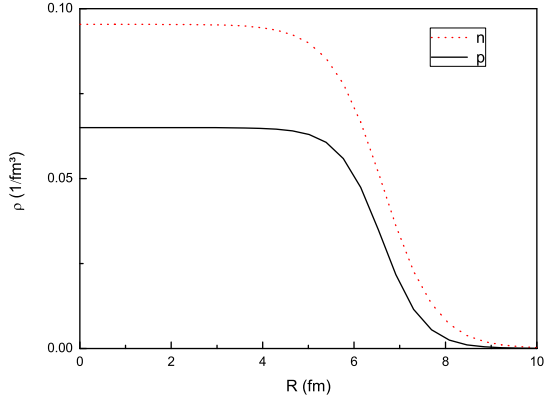}}
\vspace*{-0.5cm}
\caption{(color online) Neutron (red) and proton (black) point densities as obtained from the 
DBHF model.                          
} 
\label{two}
\end{center}
\end{figure}

\section{Symmetry pressure, neutron skin, and neutron star radii in microscopic models} 

Typically, 
predictions of the symmetry energy at saturation density encountered in the literature are in reasonable agreement with 
one another, ranging approximately from 26 to 35 MeV. On the other hand, 
the slope of the symmetry energy is more model dependent.                          
This is seen through the symmetry pressure, defined as 
\begin{equation}
L = 3 \rho_0 \Big (\frac{\partial E_{sym}(\rho)}{\partial \rho}\Big )_{\rho_0} \approx 
 3 \rho_0 \Big (\frac{\partial e_{n.m.}(\rho)}{\partial \rho}\Big )_{\rho_0}.
\end{equation} 
Thus, $L$ is sensitive to the gradient of the energy per particle in neutron matter ($e_{n.m.}$). 
 Clearly, the neutron skin, given by                                  
\begin{equation}
S = \sqrt{<r_n^2>} - \sqrt{<r_p^2>}\; , 
\end{equation} 
is highly sensitive to the same gradient.

Values of $L$ are reported to range 
from -50 to 100 MeV as seen, for instance, through the numerous
parametrizations of Skyrme interactions (see Ref.~\cite{BA} and references therein),               
 all chosen to fit the binding energies and the 
charge radii of a large number of nuclei.  
Heavy-ion data impose boundaries for $L$ at $85 \pm 25$ MeV \cite{Chen07,Dan07}. Also, a nearly linear
correlation is observed between the neutron skin $S$ and the $L$ parameter, see Fig.~3.     
(The shown correlation is taken from Ref.~\cite{Li05}.) 
More stringent constraints are being extracted \cite{Tsang}.

Such phenomenological studies are very useful, but, 
ultimately, {\it ab initio} approaches must be employed in order to get true        
physical insight. By {\it ab initio}, we mean that the starting point is a realistic
two-body potential, possibly complemented by three-body forces. 
The tight connection with the underlying two-body potential will then facilitate the physical understanding,
when combined with reliable constraints.

Our model does not include three-body forces (TBF) explicitely, but incorporates the 
class of TBF originating from the presence of nucleons and antinucleons (the          
``Z-diagrams"), which are 
effectively accounted for in the DBHF scheme \cite{Sam0807}.

As the other main input of our comparison, we will take the EoS's from the microscopic approach 
of Ref.~\cite{cat}. There (and in previous work by the authors), 
the Brueckner-Hartree-Fock (BHF) formalism is employed along with microscopic three-body forces.        
However, in Ref.~\cite{cat}
the meson-exchange TBF are constructed applying the same parameters
as used in the corresponding nucleon-nucleon (NN) potentials, which are: Argonne V18 (V18, \cite{V18}), Bonn B (BOB, \cite{Mac89}), Nijmegen 93 (N93, \cite{N93}). 
The popular (but phenomenological) Urbana TBF (UIX, \cite{UIX}) is also utilized in Ref.~\cite{cat}. Convenient
parametrizations in terms of simple analytic functions are given in all cases and we
will use those to generate the various EoS's. We will refer to this approach, generally, as ``BHF + TBF".

In Fig.~4, we display our DBHF predictions for the symmetry energy, solid black curve, 
along with those from V18, BOB, UIX, and N93. 
All values of the symmetry energy at the respective saturation densities are 
between 29 and 34 MeV. A larger spreading is seen in the $L$ parameter, 
see Fig.~5, where the values range from about 70 to 106 MeV. The respective neutron skin predictions are shown on the vertical axis.

We notice that all BHF+TBF models predict larger $L$, and thus larger neutron skins, compared to DBHF, corresponding to 
a faster growth of the energy per particle in neutron matter relative to symmetric matter.               
This can be seen in Fig.~4, especially for the higher
densities.                                                                          
The present calculations reveal that there are more subtle, but significant differences at low to medium densities as well.
We recall that, in our DBHF calculation,                                             
the growth of the  symmetry energy (especially at high density) is moderated by the repulsive
and strongly density
dependent ``Dirac" effect, which impacts all partial waves and, in
particular, the T=0 partial waves, absent in neutron matter \cite{Sam0806}.

\begin{figure}
\begin{center}
\vspace*{-1.5cm}
\hspace*{-2.0cm}
\scalebox{0.3}{\includegraphics{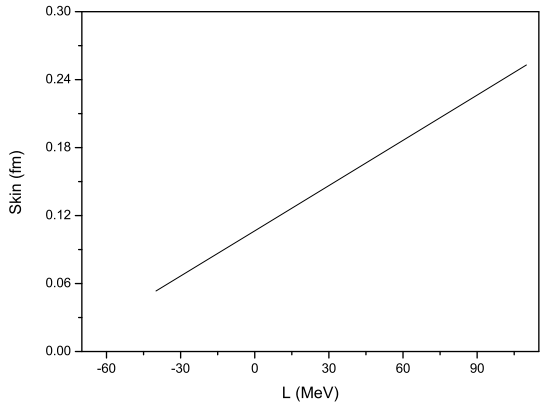}}
\vspace*{-0.5cm}
\caption{(color online) Relation between the neutron skin in $^{208}$Pb and the $L$ parameter as found in 
Ref.~\cite{Li05}.  
} 
\label{three}
\end{center}
\end{figure}

\begin{figure}
\begin{center}
\vspace*{-.5cm}
\hspace*{-2.0cm}
\scalebox{0.3}{\includegraphics{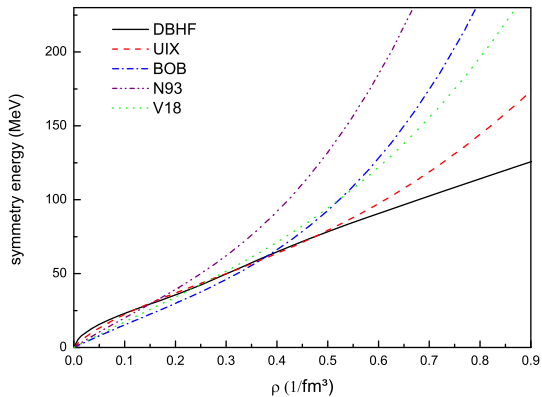}}
\vspace*{-0.5cm}
\caption{(color online) Predictions for the symmetry energy from DBHF and various
``BHF + TBF" models considered in the text. 
} 
\label{four}
\end{center}
\end{figure}

\begin{figure}
\begin{center}
\vspace*{1.0cm}
\hspace*{-2.0cm}
\scalebox{0.3}{\includegraphics{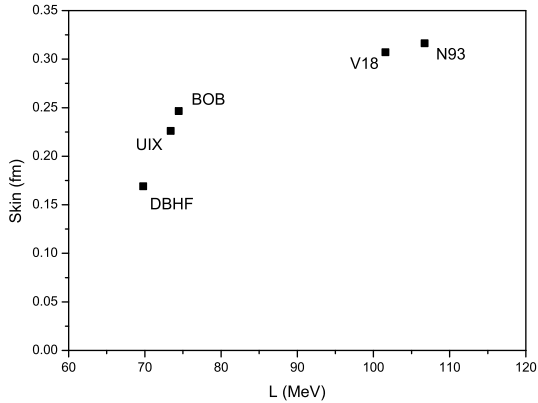}}
\vspace*{0.4cm}
\caption{(color online) Neutron skin of $^{208}$Pb {\it vs.} the symmetry pressure energy for the models
considered in the text. 
} 
\label{five}
\end{center}
\end{figure}

The next term in the expansion of the symmetry energy is the 
$K_{sym}$ parameter,
\begin{equation}
K_{sym} = 9 \rho_0^2 \Big (\frac{\partial^2 E_{sym}(\rho)}{\partial \rho ^2}\Big )_{\rho_0},  
\end{equation} 
which is a measure for the curvature of the symmetry energy. The neutron skin {\it vs.\ } $K_{sym}$ 
is shown in Fig.~6 for the various models. Although the  
values of $K_{sym}$ appear more spread out, the large negative values obtained with 
some of the parametrizations of the Skyrme model are not present.     
Those                        
large negative values (as low as -600 MeV) produced by Skyrme models indicate a strongly downward
curvature already at low to medium densities.                                          
We also notice from Fig.~6 
that no clear correlation can be identified between $K_{sym}$ and the neutron skin.    
\begin{figure}[h]
\begin{center}
\vspace*{-2.0cm}
\hspace*{-2.0cm}
\scalebox{0.3}{\includegraphics{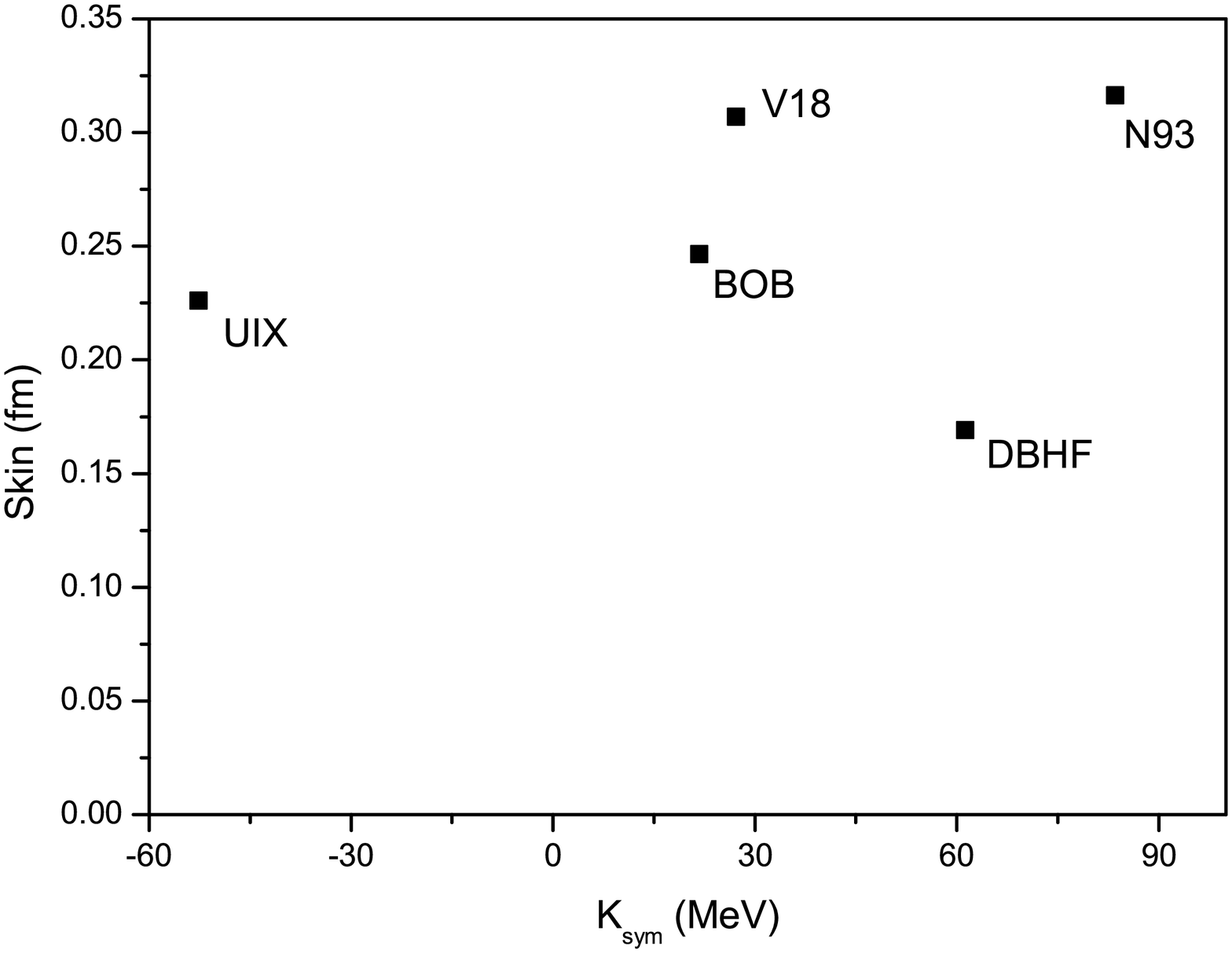}}
\vspace*{0.5cm}
\caption{(color online) Neutron skin of $^{208}$Pb {\it vs.\ } the curvature of the symmetry energy, $K_{sym}$, for the models
considered in the text. 
} 
\label{six}
\end{center}
\end{figure}

Going back to the correlation between neutron skin and symmetry pressure, we now wish to further explore the nature
of such correlation. 
For that purpose,                                                 
we made the following test: from our baseline DBHF model, we applied {\it ad hoc} 
variations of the symmetry energy {\it only}, leaving the symmetric matter EoS unaltered. Specifically,
we increased the symmetry energy in small steps, by an amount between 5\% and 20\%, and recalculated
the neutron skin of $^{208}$Pb for each of these ``varied models". (Note that the charge radius didn't change in any significant
manner as a result of these modifications.) The corresponding skin-$L$ correlation is much closer to a linear 
one, compare Fig.~5 and  Fig.~7. 
The actual models considered in Fig.~5 are quite different from one another,
in both the symmetric matter EoS and the symmetry energy. And, although the general pattern is that larger
$L$ corresponds to larger neutron skin, the actual relation is more complex, as the r.m.s. radius of the neutron 
distribution will also receive feedback from the smaller or larger degree of attraction which binds the neutrons 
to the protons and which is 
determined by the symmetric matter EoS. 
\begin{figure}[h]
\begin{center}
\vspace*{0.5cm}
\hspace*{-2.0cm}
\scalebox{0.3}{\includegraphics{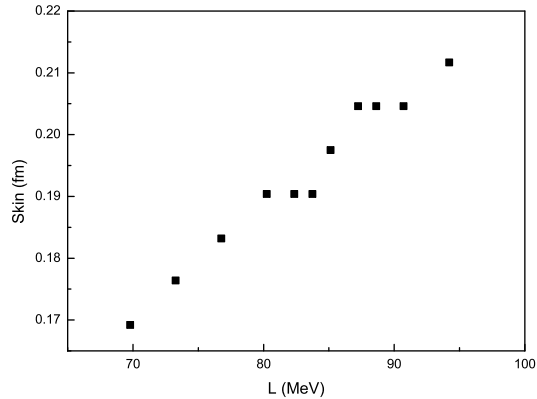}}
\vspace*{0.5cm}
\caption{(color online) The relation between the                                 
neutron skin of $^{208}$Pb and the symmetry pressure for the                                             
 DBHF model and {\it ad hoc } 
variations of the symmetry energy as explained in the text. 
} 
\label{seven}
\end{center}
\end{figure}

We conclude this section by showing in Fig.~8 the relation between the neutron skin of $^{208}$Pb and the radius of 
a 1.4M$_{\odot}$ neutron star for the models shown in Fig.~5.                                                     
(For simplicity, we consider here only pure neutron matter.) 
At first, the figure can appear surprising, since
 larger skin does not necessarily imply larger radius. On the other hand, one must keep in mind that:
1) The radius depends mostly on the pressure at the higher densities, whereas the skin probes normal or 
subnuclear densities;         
 2) As we argued in the previous paragraph for the case of a nucleus, these 
models differ from  one another in more than just the slope of the symmetry energy. 
\begin{figure}[h] 
\begin{center}
\vspace*{-2.5cm}
\hspace*{-2.0cm}
\scalebox{0.3}{\includegraphics{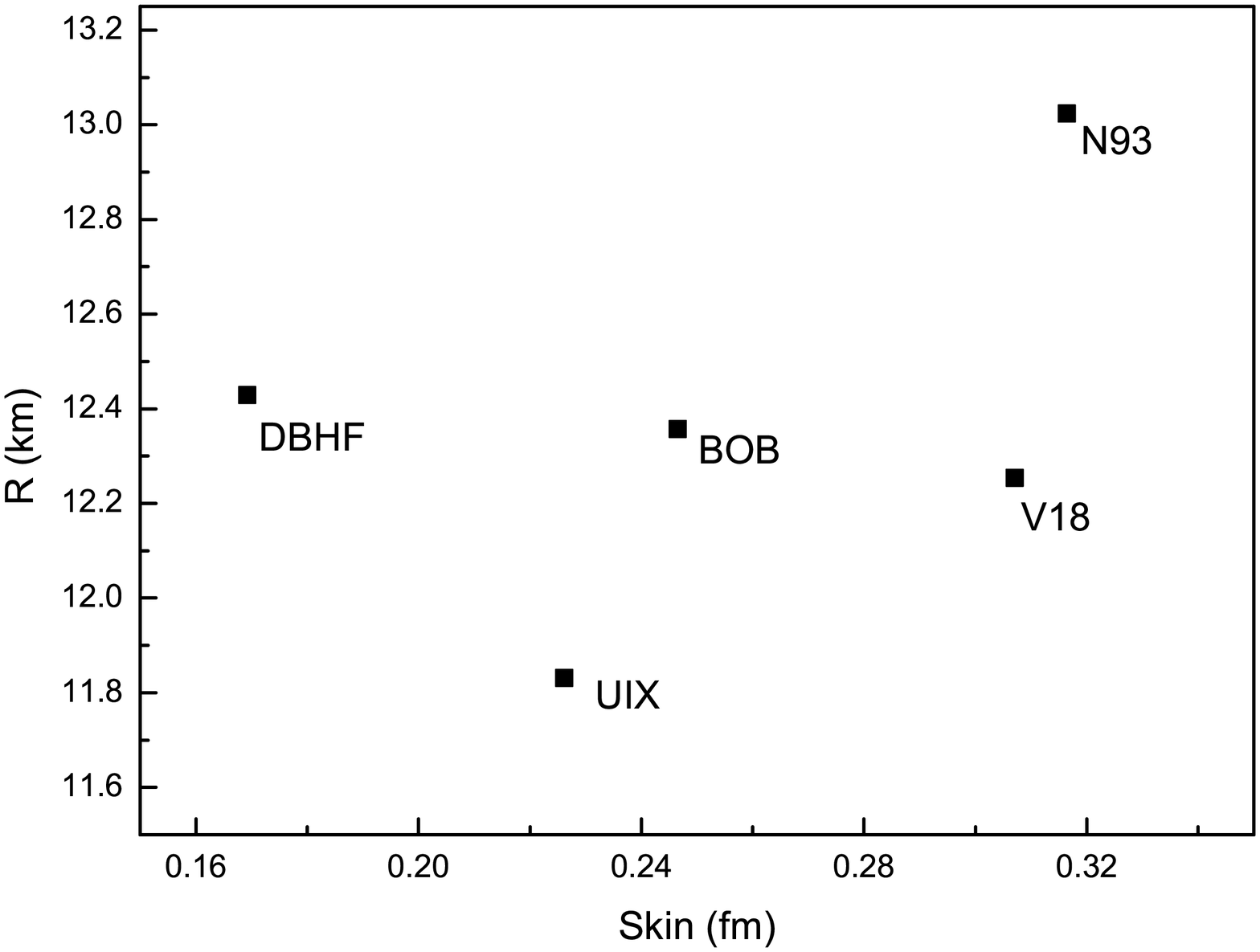}}
\vspace*{0.2cm}
\caption{(color online) Neutron skin of $^{208}$Pb {\it vs.\ } the radius of a 1.4M$_{\odot}$ neutron star for the models
considered in the text. 
} 
\label{eight}
\end{center}
\end{figure}
                        
\begin{table}[h] 
\caption{Radii and central energy densities of a 1.4M$_{\odot}$ neutron star predicted by each model. 
} 
\begin{center}
\begin{tabular}{|c|c|c|}
\hline
Model & R(km) & $e_c$ (10$^{14}$gm/cm$^3$)  \\                                                    
                  \cline{1-3}
DBHF &   12.4 & 6.5                                              \\ 
UIX &    11.8 & 9.05                                                 \\ 
V18   &  12.3 & 6.89              \\                                     
BOB &    12.4 & 6.16                                    \\ 
N93 &    13.0 & 5.91                                    \\ 
\cline{1-3}
\hline
\end{tabular}
\end{center}
\end{table}
Table I can give additional insight. Even though UIX  predicts a larger skin than DBHF, its predicted neutron star radius
is 11.8 km as opposed to 12.4 km from DBHF. Notice, however, the much larger central energy density in the UIX case. 
In other words, UIX  produces a more compact 1.4M$_{\odot}$ neutron star.
BOB and V18 are quite similar to DBHF in both radius and central energy density. 
On the other hand, in a case like N93, where, as seen from Fig.~4, high pressure is sustained pretty much 
at all densities, 
the 1.4M$_{\odot}$ star is larger and more diffuse.                                     
In summary, 
for each model the available mass will distribute itself differently, depending on both 
 the symmetric and the asymmetric part of the EoS {\it at each relevant density}. The different values of the predicted maximum masses,
apparent from Fig.~9, reflect differences among the models in their symmetric matter EoS's. 

To complete this test, we took again 
the arbitrarily varied ``models" used in Fig.~7 and reexamined the radius {\it vs.}
$L$ (or {\it vs.} neutron skin) relation, see Figs.~10 and 11. The correlations are nearly linear.  
For completeness, the mass-radius relation for these 
varied ``models" is displayed in Fig.~12, and shows that the same maximum mass 
is predicted in all cases (as expected, since the symmetric matter EoS is unchanged).

\section{Outlook}                                                                  
The neutron skin is an important isospin-sensitive ``observable" which is 
essentially determined by the difference in pressure between symmetric and 
neutron matter. Interesting correlations can be established between the skin 
of a heavy nucleus (e.~g. $^{208}$Pb) and neutron star properties 
\cite{Piek06}. 
We calculated the neutron skin of $^{208}$Pb with our latest EoS based
on  the  DBHF approach \cite{Sam0806}, as well as other EoS's based on BHF calculations
with TBF, with the parameters of the meson-exchange TBF consistent with those
of the underlying meson-exchange NN potential.

Most models agree on the value of $E_{sym}$ around the saturation point within a few MeV, 
but we are far from a reasonable agreement on the slope of the symmetry 
energy and even farther from agreement on the curvature.
Although microscopic models do not display as much spread as phenomenological ones, 
a  point that comes out clearly from the present study is that 
a measurement of the neutron skin of $^{208}$Pb with an accuracy of 0.05 fm, as it has been announced \cite{Hor}, should definitely
be able to discriminate among EoS's from microscopic models. 
\begin{figure}[h]
\begin{center}
\vspace*{1.5cm}
\hspace*{-2.0cm}
\scalebox{0.3}{\includegraphics{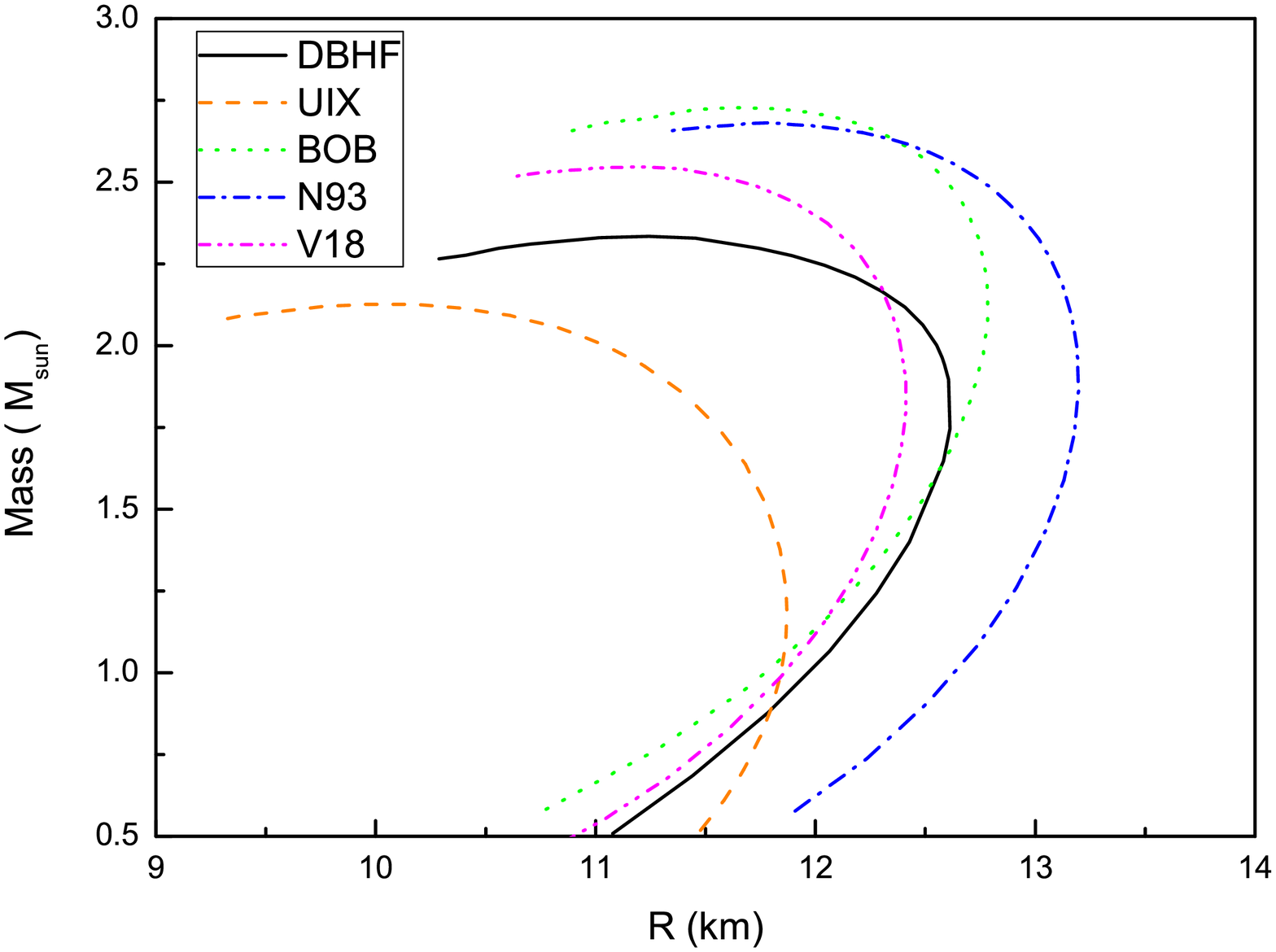}}
\vspace*{0.2cm}
\caption{(color online) Neutron star mass-radius relation for the models
considered in the text. 
} 
\label{nine}
\end{center}
\end{figure}
\begin{figure}[h]
\begin{center}
\vspace*{-1.5cm}
\hspace*{-2.0cm}
\scalebox{0.29}{\includegraphics{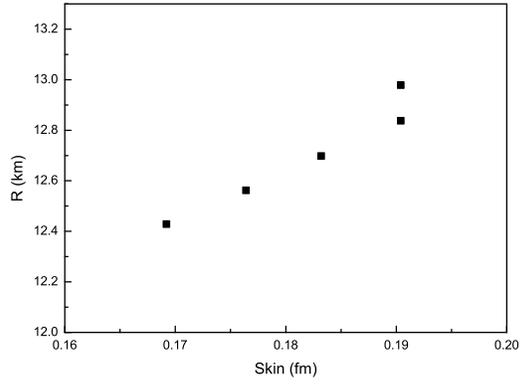}}
\vspace*{0.2cm}
\caption{(color online) Neutron skin of $^{208}$Pb {\it vs.\ } the radius of a 1.4M$_{\odot}$ neutron star for the DBHF  
model and {\it ad hoc} variations as explained in the text. 
} 
\label{ten}
\end{center}
\end{figure}
\begin{figure}[h]
\begin{center}
\vspace*{-0.5cm}
\hspace*{-2.0cm}
\scalebox{0.28}{\includegraphics{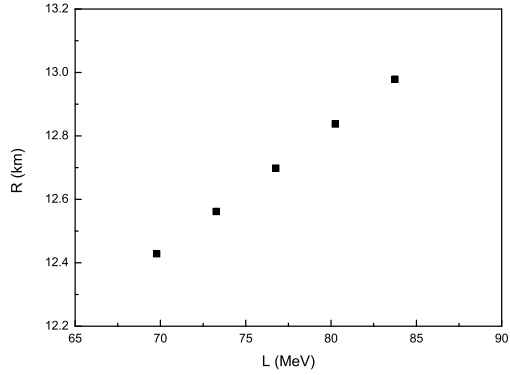}}
\vspace*{0.2cm}
\caption{(color online) The symmetry pressure, $L$, {\it vs.} the radius of a 1.4M$_{\odot}$ neutron star for the DBHF model 
and {\it ad hoc} variations as explained in the text. 
} 
\label{eleven}
\end{center}
\end{figure}
\begin{figure}[h] 
\begin{center}
\vspace*{-1.7cm} 
\hspace*{-2.0cm}
\scalebox{0.29}{\includegraphics{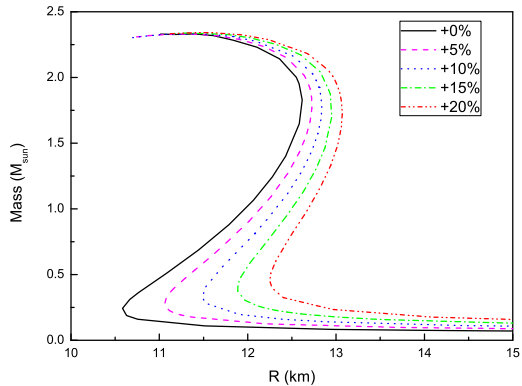}}
\vspace*{0.2cm}
\caption{(color online) Neutron star mass-radius relation for the DBHF model and {\it ad hoc}
variations as explained in the text. The legend on the upper right corner indicates the amount
by which the symmetry energy has been increased in each case. 
} 
\label{twelve}
\end{center}
\end{figure}

\section*{Acknowledgments}
Support from the U.S. Department of Energy under Grant No. DE-FG02-03ER41270 is 
acknowledged.                                                                           
%\section{References}


\newpage
\begin{thebibliography}{99}
\bibitem{Piek06} J. Piekarewicz, arXiv:nucl-th/0607039, and references therein.
\bibitem{AS03} D. Alonso and F. Sammarruca, Phys. Rev. C {\bf 68}, 054305 (2003).     
\bibitem{Mac89} R. Machleidt, Adv. Nucl. Phys. {\bf 19}, 189 (1989). 
\bibitem{Sam0806} F. Sammarruca and Pei Liu, arXiv:0806.1936 [nucl-th]. 
\bibitem{AS03I} D. Alonso and F. Sammarruca, Phys. Rev. C {\bf 67}, 054301 (2003).     
\bibitem{Oya98} K. Oyamatsu {\it et al.}, Nucl. Phys. {\bf A634}, 3 (1998). 
\bibitem{Furn} R.J. Furnstahl,             
Nucl. Phys. {\bf A706}, 85 (2002).                      
\bibitem{BA} B.A. Li and L.W. Chen, Phys. Rev. C {\bf 72}, 064611 (2005).         
\bibitem{Chen07}  L.W. Chen {\it et al.},                                                
arXiv:0711.1714 [nucl-th]. 
\bibitem{Dan07} Pawel Danielewicz and Jenny Lee, 
arXiv:0708.2830 [nucl-th]. 
\bibitem{Li05} Lie-Wen Chen, Che Ming Ko, and Bao-An Li, arXiv:nucl-th/0509009.           
\bibitem{Tsang} M.B. Tsang, Y. Zhang, P. Danielewicz, and M. Famiano, unpublished.
\bibitem{Sam0807} F. Sammarruca, arXiv:0807.0263 [nucl-th], and references therein. 
\bibitem{cat} Z.H. Li and H.-J. Schulze, Phys. Rev. C {\bf 78}, 028801 (2008).     
\bibitem{V18} R.B. Wiringa, V.G.J. Stocks, and R. Schiavilla, Phys. Rev. C {\bf 51}, 38 (1995).
\bibitem{N93} V.G.J. Stocks, R.A.M. Klomp, C.P.F. Terheggen, and J.J. de Swart, Phys. Rev. C {\bf 49}, 2950 (1994).
\bibitem{UIX} S.C. Pieper, V.R. Pandharipande, R.B. Wiringa, and J. Carlson, Phys. Rev. C {\bf 64}, 014001 (2001). 
\bibitem{Hor} C.J. Horowitz, S.J. Pollock, P.A. Souder, and R. Michaels, Phys. Rev. C {\bf 63}, 025501 (2001).     
\end{thebibliography}
\end{document}